# The role of MHD in causing impurity peaking in JET Hybrid plasmas

T C Hender[1], P Buratti[2], F J Casson[1], B Alper[1], Yu Baranov[1], M Baruzzo[3], C D Challis[1], F Koechl[4], C Marchetto[5], M F F Nave[6], T Pütterich[7], S Reyes Cortes[6], and JET Contributors*

*EUROfusion Consortium, JET, Culham Science Centre, Abingdon, OX14 3DB, UK*

[1]*CCFE, Culham Science Centre, Abingdon, OX14 3DB, UK*

[2]*ENEA for EUROfusion Via E. Fermi 45, 00044 Frascati (Roma),*

[3]*RFX, Corso Stati Uniti 4, Padova, Italy,*

[4]*ÖAW/ATI, Atominstitut, TU Wien, 1020 Vienna, Austria*

[5]*Istituto di Fisica del Plasma, CNR, Milano, Italy*

[6]*Instituto de Plasmas e Fusão Nuclear, Instituto Superior Técnico, Universidade de Lisboa1049-001, Lisboa, Portugal*

[7]*Max-Planck-Institut für Plasmaphysik, D-85748 Garching, Germany,*

*See the Appendix of F. Romanelli et al, Proc. of the 25th IAEA FEC, Saint Petersburg, Russia, 2014

**Abstract**

In Hybrid plasma operation in JET with its ITER-like wall (JET-ILW) it is found that $n>1$ tearing activity can significantly enhance the rate of on-axis peaking of tungsten impurities, which in turn significantly degrades discharge performance. Core $n=1$ instabilities can be beneficial in removing tungsten impurities from the plasma core (e.g. sawteeth or fishbones), but can conversely also degrade core confinement (particularly in combination with simultaneous $n=3$ activity). The nature of MHD instabilities in JET Hybrid discharges, with both its previous Carbon wall and subsequent JET-ILW, is surveyed statistically and the character of the instabilities is examined. Possible qualitative models for how the $n>1$ islands can enhance on-axis tungsten transport accumulation processes are presented.

**1 Introduction**

A promising regime for long pulse operation in ITER is the so-called Hybrid regime. In this regime tailoring of the current profile is employed to give low magnetic shear in the core region, with a central safety factor $q(0)\sim1$. The current profile tailoring may be in terms of how the additional heating is sequenced and/or how plasma current is ramped from discharge initiation. The Hybrid regime has been explored on DIII-D [1, 2], ASDEX-Upgrade [3, 4], JT60-U [5] and can show enhanced energy confinement compared to the baseline ELMy H-mode scaling [6]; with confinement enhancement factors $H_{98y2}>1$. This improved energy confinement seems linked with the achievement of higher $\beta$ in the Hybrid regime [7].

Hybrid plasmas has also been explored on JET; initially these studies started in the carbon wall (JET-C) phase of JET operation [8, 9], but have continued in the ITER-like Wall (JET-ILW) phase of JET operation [7]. In the JET-ILW phase the main plasma facing wall is beryllium coated, but the power handling elements of the divertor are tungsten, mimicking the first wall materials in ITER. Since tungsten is a high-Z material it can radiate significantly, introducing an additional performance limitation mechanism compared to JET-C operation in JET – this is particularly problematic if the tungsten accumulates in the plasma core. In some cases there seems to be an interplay between low toroidal mode number ($n$) instabilities in JET-ILW Hybrid plasmas and the occurrence of core impurity accumulation [10], that leads to significant performance degradation. An example of such behaviour is shown in Figs. 1 and 2 for a JET-ILW Hybrid plasma pulse.

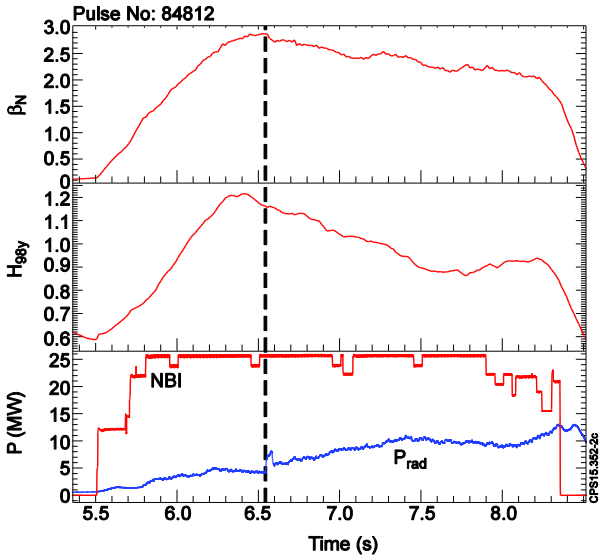

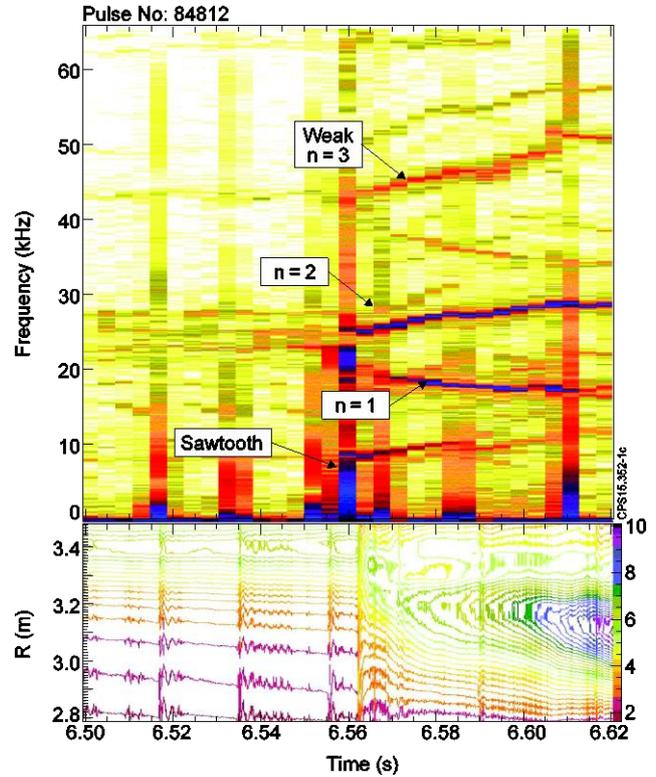

**Fig 1** Normalised $\beta_N$, confinement H-factor, radiated power ($P_{rad}$) and NBI power. Here $\beta_N$ =($\beta_t/(I_p(MA)/B(T) a(m))$), with $a(m)$ the plasma minor radius and $B(T)$ the vacuum toroidal field. The start timing of the n=2 mode is indicated by the broken vertical line.

**Fig 2** *Example of an m=3, n=2 tearing mode causing rapid impurity peaking in a JET-ILW Hybrid pulse with $I_p$=2.0MA , $B_t$ =2.42T ($q_{95}$=3.95). The upper plot shows the spectrogram of an outboard coil measuring poloidal magnetic fluctuations. The lower contour plot is of SXR emission from an array of vertically viewing SXR channels, where their major radius (R) is determined from their intercept with the plasma mid-plane.*

From Fig 1 it can be seen that the confinement goes into sharp decline at the time an $n$=2 mode starts (indicated by the vertical broken line). Figure 2 shows the soft X-ray (SXR) emission is dominantly peaked off-axis ($R$~3.4m) until the initiation of the $m$=3, $n$=2 mode at 6.56s (here $m$ is the poloidal mode number, identified from the radial location of the mode). This peaking corresponds to off-axis localisation of the tungsten on the low field side (LFS) due to inward radial neo-classical transport, with its LFS localisation due to centrifugal effects in the parallel force balance [11]. As the $n$=2 mode onsets it can be seen that the SXR emission (from tungsten predominantly) rapidly peaks on-axis ($R$~3.15m). As a result of the combined effect expected from the $n$=2 mode and the strong core radiation there is a substantial confinement degradation (see Fig 1). The relationship of MHD, tungsten transport and resultant effect on discharge performance is the subject of this paper. Since relatively equivalent Hybrid discharges were studied in the JET-C and JET-ILW phases of operation, it is helpful to compare the effects of MHD instabilities in these 2 cases – this is done in this paper. With the JET-C previous work has shown that $n$=1 and 2 modes can affect the transport of argon and neon [12].

Previously it has been shown $n$>1 tearing instabilities only accelerate the on-axis tungsten impurity accumulation if the impurity is localised off-axis – if it is already peaked on-axis then the instability has no significant effect on the peaking [11]. The role of the MHD in some discharges is to accelerate the on-axis peaking, but in many discharges peaking occurs without MHD playing a major role [11].

In the remainder of this paper we discuss for a database of JET-C and JET-ILW Hybrid pulses, the types of low-$n$ MHD observed and its effects (section 2), and then examine in more detail the structure and effects of the observed $n$>1 instabilities (section 3) and $n$=1 instabilities (section 4) – in all cases the commonality and differences between the JET-ILW and JET-C are examined. In section 5 possible models for the interaction of MHD instabilities and high-Z impurity transport are examined, and finally in section 6 conclusions are given.

## 2  Statistical Survey

A database of all 262 Hybrid pulses in the JET-ILW phase of operation and a representative set of 181 JET-C Hybrid pulses has been considered. MHD instabilities are classified in discharges for which at least 5MW of neutral beam injection (NBI) power is applied and for which the amplitude the poloidal magnetic field fluctuations must for >0.1s exceed $10^{-4}$T for $n$=1, $3\times10^{-5}$T for $n$=2 and $10^{-5}$T for $n$=3. This toroidal mode number analysis is performed using 1MHz bandwidth pick-up coils located slightly above the mid-plane on the outboard side of the JET vessel and the quoted mode amplitudes are those observed at the coils. Also a radiation criteria is applied, that modes appearing after $P_{rad}/P_{input}$ exceeds 70% are discarded - this eliminates pulses which are dominated by early impurity influxes and hollow temperature profiles (here $P_{rad}$ is the total radiated power and $P_{input}$ the total additional heating power from neutral beam and ion cyclotron(if present)). In Fig 3 the number of pulses with $n$=1, 2 or 3 MHD are shown for the JET-ILW and JET-C datasets. The $n$=2 activity is identified from its location to be occurring at the $q$=3/2 surface – this is done by matching the MHD activity frequency against the rotation profile measured by charge exchange [13]. Generally the $n$=3 activity is at the $q$=4/3 surface, but in some pulses it is at the 5/3 surface. In some pulses the main MHD activity is $m$=2, $n$=1 (as opposed to core $m$=$n$=1) and the figures in the semi-annular region record the occurrence of these pulses. In this case the $m$=2, $n$=1 is discriminated from $m$=$n$=1 activity by its mode location.

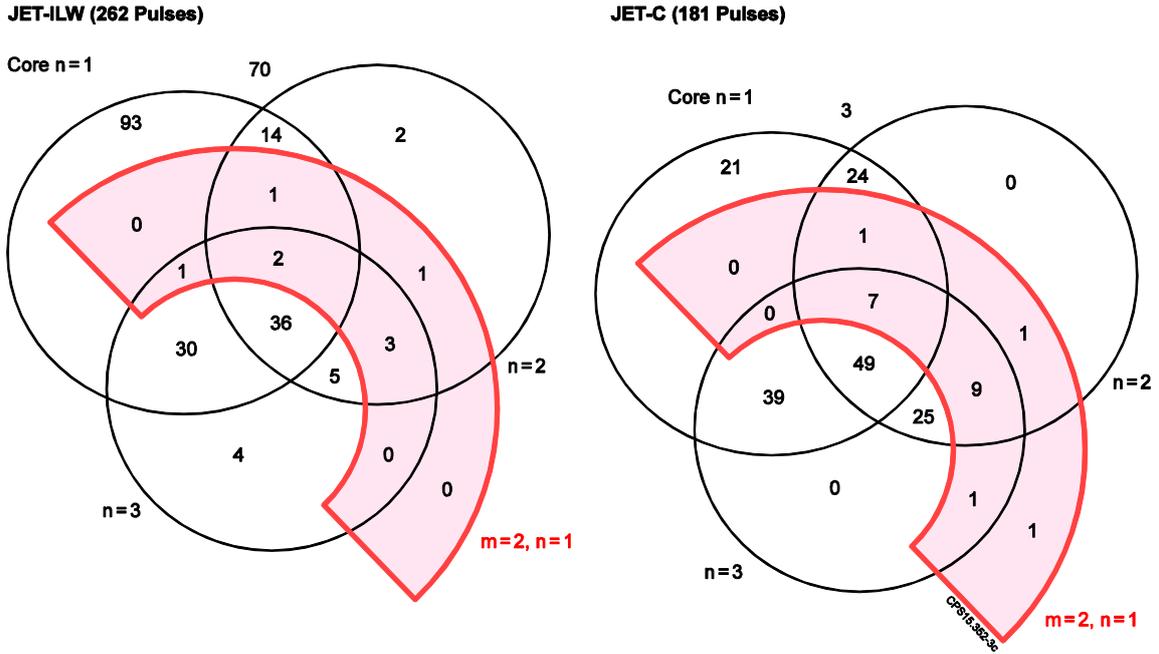

**Fig 3** *Venn diagram of core n=1, (m=2, n=1), n=2 or n=3 MHD in JET-ILW and JET-C pulses. The numbers outside the circles are those without MHD (e.g. 70 pulses for the JET-ILW)*

From Fig 3 it can be seen that the majority of pulses with MHD activity present have core *n*=1 activity. With the JET-C *n*=2 or *n*=3 MHD activity is more common and correspondingly *n*=1 activity alone, is much less common. Some insight into the reasons for the differences in observed modes between the JET-C and JET-ILW can be found by examining the domain of their occurrence (Fig 4).

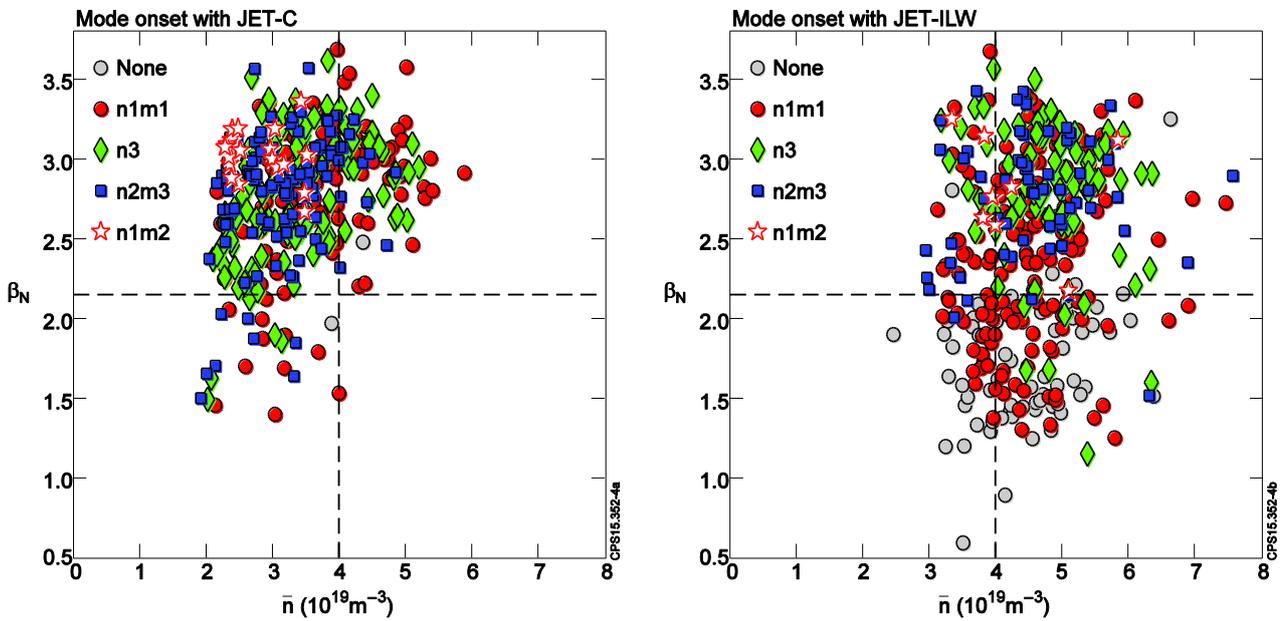

**Fig 4** *Occurrence of core n=1, n=2 or n=3, no MHD, and m=2 n=1, as a function of $\beta_N$ and line averaged electron plasma density. The broken line cross-hairs are to help guide the eye between the 2 plots.*

From Fig 4 it can be seen that the proportion of pulses with lower $\beta_N$ is much higher with the JET-ILW; over the datasets the average maximum value of $\beta_N$ is 2.44 and 3.03 for the JET-ILW and JET-C, respectively. Since, as will be shown later, the *n*=2 and 3 modes are identified as at least in-part pressure driven by neo-classical effects, the lower $\beta_N$ makes them less likely on average to be destabilised in JET-ILW Hybrid pulses (and conversely *n*=1 only instabilities, or pulses with no classified MHD, are more likely to occur). In the JET-C Hybrid pulses the domain with $\beta_N$<2.2 and $n_e$>4x10$^{19}$m$^{-3}$, where a lot of the *n*=1 only or no MHD pulses occur, was not accessed. With the JET-ILW higher density operation is needed to avoid impurity influx.

The duration of the high performance phases of Hybrid pulses before MHD onsets, is in general longer with the JET-C than the JET-ILW (Fig 5).

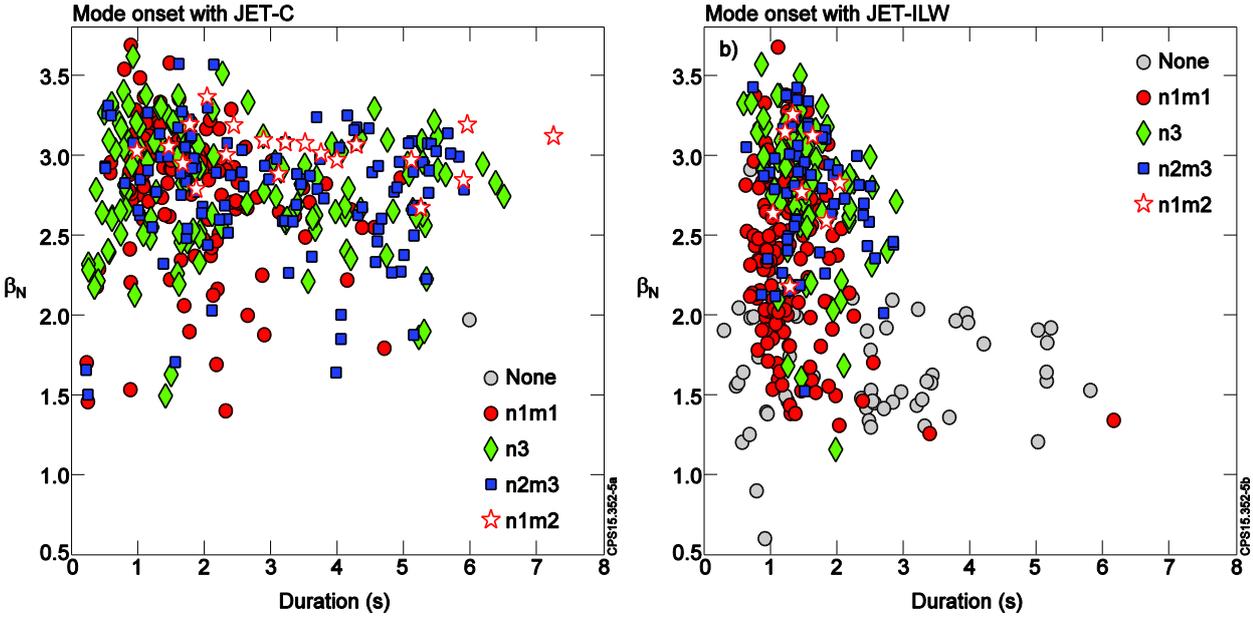

**Fig 5** *Occurrence of core n=1, n=2 or n=3, no MHD, and m=2, n=1, as a function of $\beta_N$ and time from the start of NBI heating to the ascribed mode onset.*

It can be seen so far that the only route to avoiding MHD for >3s, with the JET-ILW, is to operate at low $\beta_N$ (<2). With the JET-ILW the domain with long heating durations and high $\beta_N$ has not been explored experimentally – for $\beta_N$ >3 the longest duration high power NBI duration is <4s (meaning the upper right quadrant in Fig 5 JET-ILW is not explored in the data). This limitation on the NBI duration is largely due to constraints on the divertor power handling in JET-ILW operation. However, it should be noted that MHD activity has already started by 2s in all JET-ILW Hybrid discharges with $\beta_N$ >3. Further, in such discharges the average $H_{98y2}$ confinement enhancement factor [6] at the time of peak $\beta_N$ is 1.21, but this has declined to 0.924 by the time the high power NBI phase ends. Examination shows this decline is often due to the MHD activity, but can also be related to impurity influx not related to MHD.

As shown in Fig 1 the *n*=2 mode significantly degrades confinement. The degradation is on average worse in the JET-ILW, Fig 6. In this figure only pulses with *n*=2 amplitude > 0.3G for at least 0.2s are considered and at the time when the amplitude exceeds 0.3G a confinement H-factor > 1.1, relative to the ITER-98y2 scaling [6], and an NBI power >13MW is required to ensure high performance Hybrid shots are being considered. The relative decrease in $\beta_N$ from the *n*=2 onset is determined 0.75s later, or at the time when the NBI power drops by 20%, if sooner. If the radiated fraction exceeds 70% the pulse is not included - this avoids other factors causing the confinement change. To quantify the *n*=2 mode amplitude, its time integral is used. It can be seen with the JET-C that the *n*=2 mode leads to a confinement reduction which is weakly correlated with the integrated mode amplitude; on average the confinement reduction ~12%. This reduction is consistent with the typical island widths [14] – a perturbation of $B_\theta(n=2)$=1.5G (or $\int B_\theta dt$=1.2Gs over 0.75s) measured at the wall corresponds to a ~10% $m$=3, $n$=2 island width. The confinement reduction can be higher with the JET-ILW due to the effect of increased core radiation.

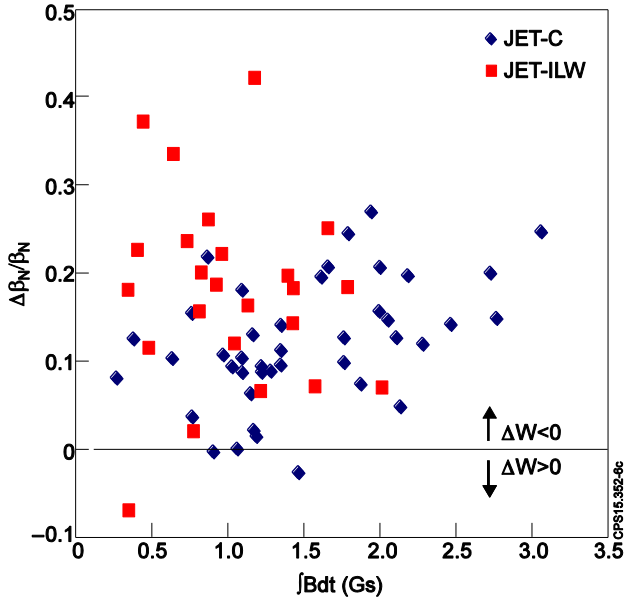

**Fig 6** *Effect of n=2 instability on confinement, the relative change in $\beta_N$ from the start of the n=2 mode to 0.75s later is shown (so that a positive value of $\Delta\beta_N$ represents a decrease in stored energy ($\Delta W$) with time), versus the integrated n=2 mode amplitude.*

In the highest performing JET-ILW pulses large $n$=1 modes can also lead to a significant confinement degradation, but this is less of an effect with the JET-C (Fig 7). In this figure only pulses with $n$=1 amplitude > 1G for at least 0.2s are considered and at the time when the amplitude exceeds 1G a confinement H-factor > 1.1, relative to the ITER-98y2 scaling [6], and an NBI power >13MW is required. The decrease in $H_{98y2}$ from the $n$=1 onset is determined 0.75s later, or at the time when the NBI power drops by 20%, if sooner. If an $n$=2 instability exceeding the threshold (0.3G) is detected during this time the shot is not included and similarly if the radiated fraction exceeds 70% it is not included - this avoids other factors causing the confinement change. To quantify the $n$=1 mode amplitude its time integral is used. The $n$=3 instability often occurs in conjunction with the $n$=1 mode (see Fig 3 Venn diagram). Those cases where the $n$=3 instability accompanies the $n$=1 mode are discriminated in Fig 7, and can be seen to lead to the largest degradation in confinement during JET-ILW operation. Apart from this case (JET-ILW and $n$=3) there is on average a small effect on confinement from the $n$=1 mode. This is particularly so for the JET-C pulses where in many cases the confinement is unaffected or improves slightly. It should be noted that there is evidence of a small, but systematic, difference in the $q$-profile between Hybrid operation with the JET-C and JET-ILW [15].

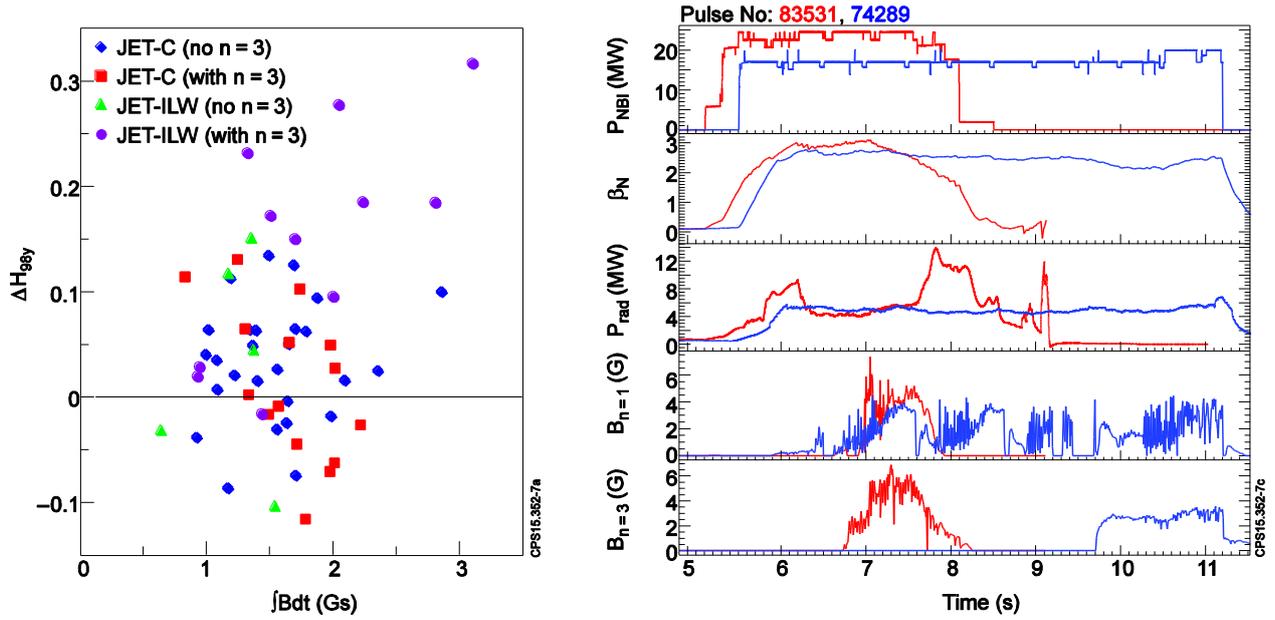

**Fig 7** *Effect of n=1 mode on confinement for the JET-ILW and JET-C. Left plot gives results using selection criteria discussed in the paper text, and shows the change in confinement ($\Delta H_{98y}$>0 represents a decrease in confinement) with integrated n=1 mode amplitude. Right plot shows typical JET-ILW (red traces) and JET-C (blue traces) pulses, with concurrent n=3 activity. Plotted are the NBI power ($P_{NBI}$), normalised pressure ($\beta_N$), total radiated power ($P_{rad}$) and the n=1 and n=3 mode amplitudes ($B_{n=1}$ and $B_{n=3}$, respectively). The JET-C pulse has sawtooth activity commencing at 6.5s, but the JET-ILW pulse has no evident sawtooth activity.*

The majority of high performance long duration (>3s at $\beta_N$>3) JET-C pulses have *n*=1 activity prior to the *n*>1 activity onset (shown in Fig 5). However the *n*=1 activity alone does not necessarily degrade confinement for the JET-C.

The domains of occurrence and global effects of the *n*=1, *n*=2 and *n*=3 MHD have been examined in this section. In the next 2 sections we look in more detail at the specific features of each type of instability.

### 3. Nature and Effects of *n*=2 modes

The structure of the mode, tearing or kink parity, can be determined by a cross phase analysis between the magnetic Mirnov signal and the 96 channels of the JET Electron Cyclotron Emission (ECE) radiometer [16]. With this technique, in regions of high coherence, islands can be discriminated by a $\pi$ phase change between neighbouring radial ECE channels. In cases identified as having coherent *n*=2 activity this technique identifies an island structure generally starting soon after the *n*=2 activity commences (though in a few cases it can take up to 0.5s for the tearing structure to become fully developed). The island is generally located close to the *q*=3/2 surface indicating an *m*=3, *n*=2 tearing mode – an example is shown in Fig 8. As mentioned above the island location can also be determined by matching the mode frequency to the measured rotation profile – this technique also gives reasonable agreement on the mode location, as shown in Fig 8. Analysis over a large set of *m*=3, *n*=2 modes shows a 1-sigma standard deviation of 3.8cm, between the *q*=1.5 location determined from the mode frequency technique versus equilibrium reconstruction including a motional Stark effect constraint [17].

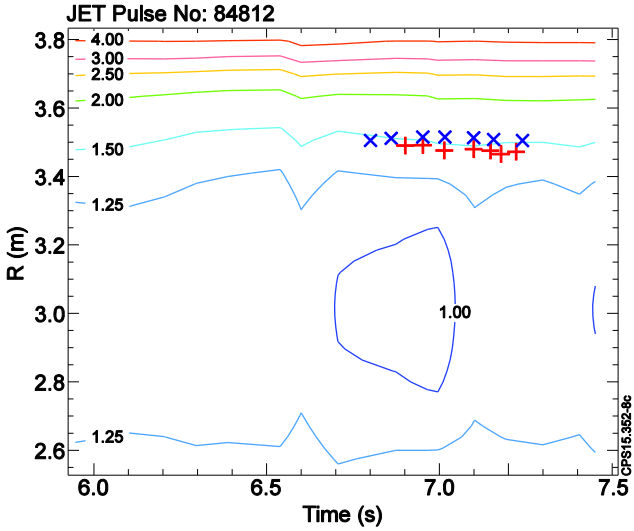

**Fig 8** *Contours of constant-q determined by the EFIT code [18] including motional Stark effect measurements as a constraint, compared with the n=2 island location from cross correlation analysis (red '+') and from its rotation frequency (blue 'x').*

Analysis shows that at the highest $\beta_N$ (>~3) the $m$=3, $n$=2 NTMs are intrinsically unstable, whereas at lower $\beta_N$ they are often sawtooth triggered [19]. This indicates that the drive for $n$=2 instabilities is a mixture of classical ($\Delta'$) effects and neo-classical drive. Indeed as the NBI power is reduced at the end of the pulse the $n$=2 island width approximately tracks $\beta_N$ (Fig 9), as would be expected for a neo-classical tearing mode (NTM).

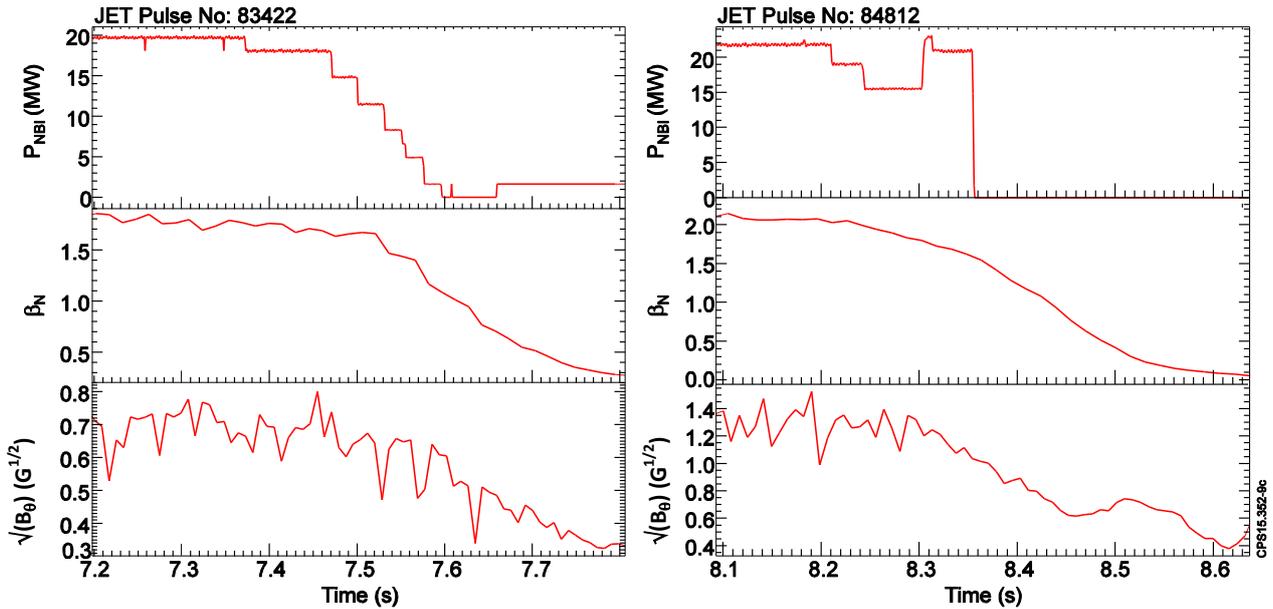

**Fig 9** *The injected NBI power, $\beta_N$, and the $\sqrt{}$[n=2 mode amplitude] (used as a proxy for the n=2 island width) for shots 83422 (left) and 84812 (right)*

The key difference between the effects of the $n$=2 instability in JET-C versus JET-ILW operation, relates to the impurity accumulation that the mode can cause with the JET-ILW. With the JET-C $n$=2 modes were quite common in Hybrid pulses (see Fig 3), but their effects were limited to a modest reduction in confinement (see Fig 6 and Fig 10) as is typical for $m$=3, $n$=2 NTMs [20]. However, with the JET-ILW provided the tungsten impurities are peaked off-axis then the $n$=2 mode rapidly accelerates the on-axis peaking. This leads to substantially increased core radiation and a much larger confinement degradation.

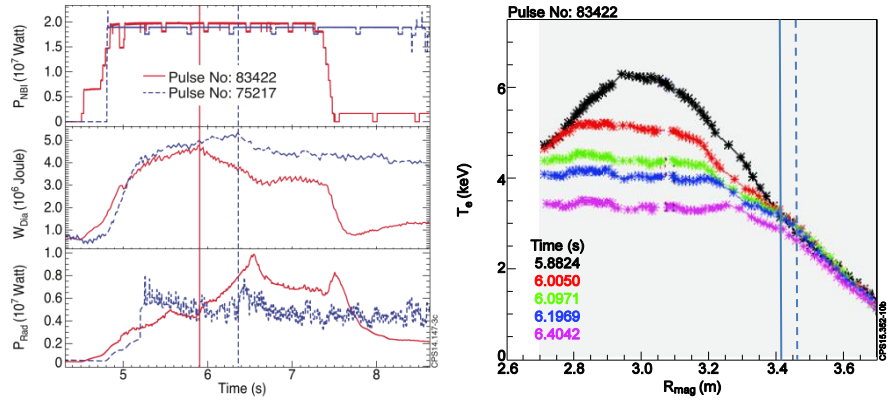

**Fig 10** *Left: Comparison of m=3, n=2 effect in JET-ILW (pulse 83422; $I_p$=1.7MA, $B_t$=2.26T) and JET-C (pulse 75217; $I_p$=1.7MA, $B_t$=1.98T). The vertical lines indicate the start time of the n=2 mode. Right: Electron temperature ($T_e$) profiles determined from Thomson scattering measurements, showing the strong degradation of $T_e$ within the m=3, n=2 mode radius. The vertical lines indicate the 3/2 radius determined from ECE cross correlation analysis (solid line) and the mode frequency (broken line).*

The location of the *m*=3, *n*=2 island compared to the off-axis peak of the tungsten is significant, in determining the mechanism by which the instability can accelerate the impurity peaking. The tungsten impurity density has been determined from analysis of 2D soft X-ray data [21, 22]. In cases where an *n*=2 mode occurs while the tungsten is still localised on the off-axis LFS, then the results show that the tungsten density tends to be radially localised just outside the radius at which the *n*=2 island forms (Fig 11).

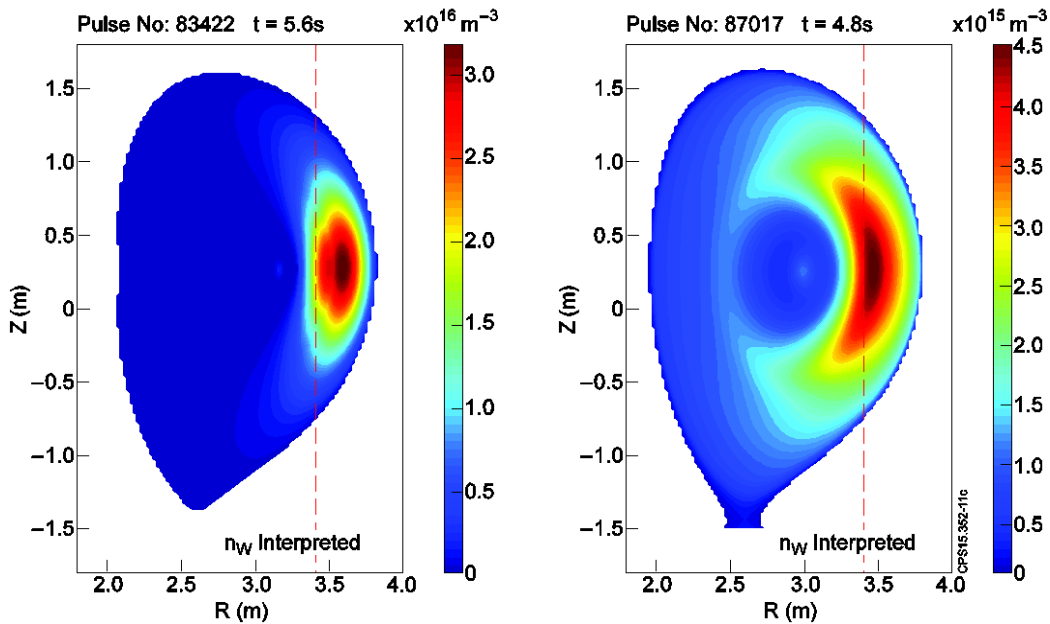

**Fig 11** *Tungsten density at times just before an n=2 mode starts. The broken line shows the radius of the phase inversion from the n=2 mode (using cross spectral analysis, as discussed above).*

As discussed above the *n*=3 modes almost always occur concurrently with *n*=1 modes during JET-ILW operation (see Fig 3) and such concurrent MHD activity has the most deleterious effect on confinement (see Fig 7).

Examination of the *n*=3 modes in such cases shows they have a tearing characteristic, though their weaker nature makes this harder to detect.

**4. Effects of core n=1 modes**

Core *n*=1 activity is by far the most common with both the JET-ILW and JET-C (see Fig 3). The effects of the core *n*=1 activity are mixed. Sawteeth are beneficial in the JET-ILW in flushing out the core impurities (Fig 12). Previously with the JET-C sawteeth have also been observed to inhibit the accumulation of argon that was introduced to produce a radiating mantle [23].

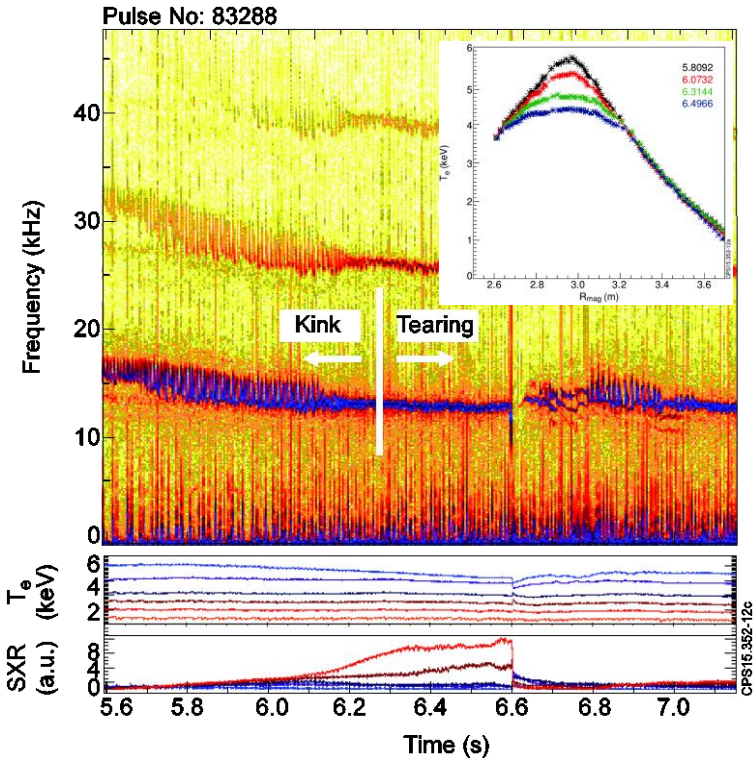

**Fig 12** *JET-ILW pulse in which peaking of the impurities in the core (indicated by the rise of the central SXR channel) is terminated by a sawtooth at ~6.6s. Note that the peaking and coherent n=1 activity is coincident. The time at which cross phase analysis shows the n=1 mode at ~15kHz changes from kink to tearing parity is indicated. The inset shows the electron temperature profile from its peak value during the fishbone activity to just before the sawtooth at 6.6s.*

From Fig 12 it can also be seen that the impurities start to peak during the period of coherent MHD after the fishbones cease – consistent with previously reported results on how fishbones inhibit (or reduce) core impurity peaking [21]. However, the confinement degradation starts during the fishbone period resulting in a reduction in central temperature and temperature gradients (Fig 12 inset). It should be noted that as the instability parity changes from kink to tearing (see Fig 12), there is no discernible effect on impurity peaking. Here the instability parity was evaluated by cross phase analysis with the ECE. The impurity peaking is likely assisted by the reduction in central temperature gradients, which reduces the outward neo-classical impurity flux term due to the temperature gradient [24]. This impurity peaking leads to an increase in core radiated power to a level that is comparable to core NBI input power density, resulting in the collapse of the central temperature (Fig 12 inset). During the phase when the core *n*=1 activity has a tearing parity, then the SXR have a helical character in the plasma core [25], which is thought be associated with accumulation of the impurities in the *m*=*n*=1 magnetic island. Observations of helical accumulation of tungsten in plasmas with *m*=*n*=1 instabilities have also been made on the ASDEX Upgrade tokamak [26]. It is known that such impurity accumulation in islands can act as an instability drive term [27].

As noted fishbones have been observed to have a beneficial effect in reducing impurity peaking [21]. With the JET-C long duration Hybrid pulses were established where the *n*=1 activity had a fishbone character and the pulses were typically only degraded late-on by *n*>1 activity – an example is shown in Fig 13.

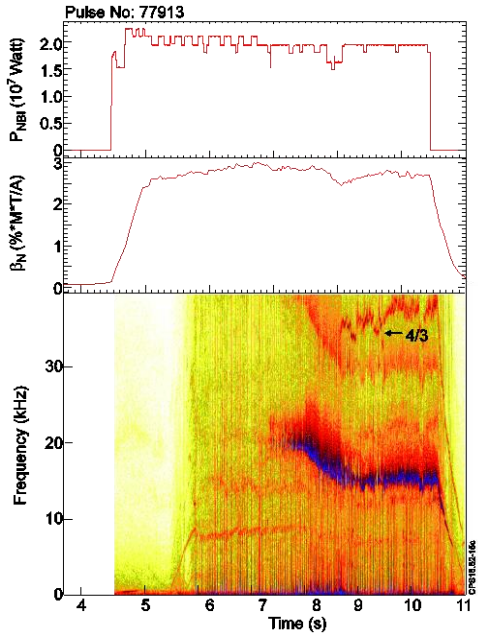

**Fig 13** *High performance Hybrid JET-C pulse. Shown are the NBI power ($P_{NBI}$) the normalised $\beta_N$ and a spectrogram of an outboard coil measuring poloidal magnetic fluctuations. The n=1 fishbone activity at ~15kHz has no major effect on confinement. The activity at >30kHz is identified as an m=4, n=3 instability.*

As discussed above in high performance JET-ILW operation, with $\beta_N$ >3, the high power NBI duration is less than 4s in Hybrid pulses, so long duration pulses of the type shown in Fig 13 have not been attempted with the JET-ILW.

## 5. Models for interaction of *n*=2 tearing modes and W transport

Two possibilities for how the *n*=2 island affects impurity transport have been considered:-

1. The island, which tends to occur inboard of the initial off-axis tungsten peak (Fig 11) increases the inward diffusion of tungsten due to parallel transport along the perturbed field lines.
2. The island affects the background deuterium fuel ion profiles, which in-turn alters neo-classical and turbulent transport of the tungsten impurity.

An initial qualitative examination of these 2 possible models is presented in this section.

### 5.1 Increased tungsten diffusion due to the island.

The idea of increasing the diffusion in the region of the island to mimic its effect on transport has been previously validated with the JETTO [28] transport code, in the case of predicting the effects of islands on neutron yield in JET-ILW Hybrid pulses [29]. In this paper we present an initial analysis of the effect of a (3,2) island on the transport of tungsten localised off-axis on the low field side. The JETTO transport code is used in interpretive mode with stationary fuel ion profiles, combined with the SANCO impurity transport code [30] used in predictive mode, with transport coefficients taken from first-principle modelling of a similar Hybrid shot [11] to obtain a hollow SXR emissivity profile with off-axis impurity localisation, that is qualitatively similar to the experiment. The impurity diffusion is then significantly increased in a position and with a width experimentally determined for the shot under examination [16]. It should be noted that the rate of rise of the *n*=2 mode amplitude, observed experimentally, is fast compared with the transport timescales and so the island effect is introduced instantaneously in the modelling. It is found that the increased island diffusion has a significant effect in peaking the tungsten in the core when the island is on the steep gradient portion of the initial off-axis impurity peak (Fig 14). The magnitude of the enhancement to the tungsten diffusion, ($D_W$), is ad-hoc at present. Decreasing the enhancement by an order of magnitude, to peak $D_W \sim 40 m^2 s^{-1}$, still produces significant effects, but a further order of magnitude reduction negates any effects from the island. Conversely increasing $D_W$ further (beyond the enhancement shown in Fig 14) does not significantly affect the predicted W-density profiles. The initial rate of rise of the central SXR channels, predicted by the model due to the W-peaking, is comparable with that observed but at this stage this model does not represent a fully quantitative comparison with experiment - future work will concentrate on this, as well as establishing the validity of the assumption of increased W radial transport, due to the island.

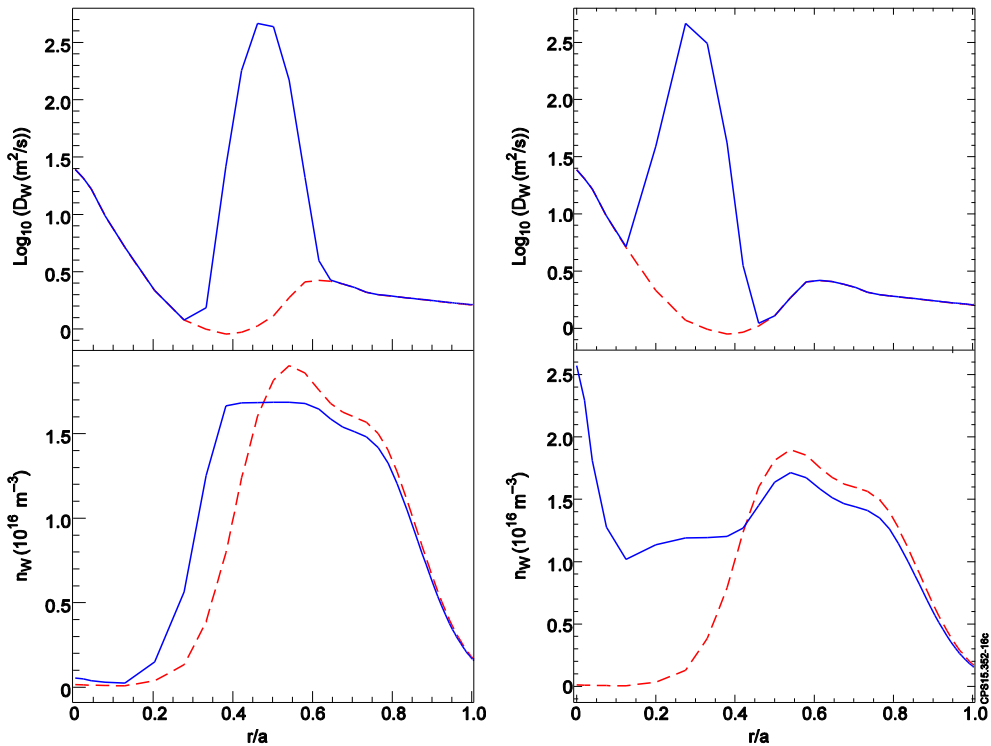

**Fig 14** *Upper row:- initial tungsten diffusion profiles (broken red line) and locally enhanced diffusion (solid blue line) for 2 different locations of the enhancement. Lower row:- shows the resultant change in the steady state tungsten impurity density ($n_W$).*

**5.2 Tungsten transport changes induced by profile changes associated with an island**

In this sub-section the alternate explanation that the island affects the fuel ion profiles, which in-turn affects the tungsten transport, is examined. The neoclassical and turbulent W transport are predicted using first-principle based modelling codes which include the poloidal asymmetries driven by rotation, and have previously been validated for JET Hybrid pulses without any NTM [11]. No specific transport due to the island is included, but the modelling analyses the *indirect* effect of the island on W transport, through its effect on the plasma profiles of deuterium fuel ions. In this model, W accumulation is driven by neoclassical convection strongly enhanced by poloidal asymmetries [31]. The direction of the neoclassical impurity convection is, as usual, determined by a competition between the main ion temperature and density gradients ($V_W \sim Z[R/L_{ni} - 0.5R/L_{Ti}]$), where $L_{ni}=n_i/(dn_i/dr)$ is the density gradient scale length and $L_{Ti}$ is the equivalently defined ion temperature scale length.

The modelling methodology is the same as described in Refs. [11,32]; the predictions of the drift-kinetic neoclassical code NEO [33, 34] are combined additively with quasilinear gyrokinetic simulations from GKW [35], normalised to match the ion heat diffusivity calculated from interpretive power balance $\chi_{an}$. Since there is no W source in the core, the ratio of total convection to diffusion components predicts a steady-state W gradient:-

$$\frac{R}{L_{n_Z}} = -\frac{\frac{\chi_{i\,an}}{\chi_{i\,NEO}} \cdot \frac{RV_{Z\,GKW}}{\chi_{i\,GKW}} + \frac{RV_{Z\,NEO}}{\chi_{i\,NEO}}}{\frac{\chi_{i\,an}}{\chi_{i\,NEO}} \cdot \frac{D_{Z\,GKW}}{\chi_{i\,GKW}} + \frac{D_{Z\,NEO}}{\chi_{i\,NEO}}}.$$

(1)

The modelling inputs are bulk plasma profiles and equilibrium geometry, based on JET-ILW Hybrid pulse 84812, in which a large $m$=3, $n$=2 mode appears at 6.55s, correlated in time with a rapid increase of central radiation (see Fig 1). At the appearance of the $n$=2 mode, the electron temperature profile exhibits a significant flattening, but the density profile does not flatten significantly, in common with other observations, and expectations from theory [36]. The fitted profiles, $T_e$, $n_e$, are produced from a single time integration of the High Resolution Thompson Scattering system (used here for its temporal resolution), with a similar $T_e$ flattening seen also using LIDAR or ECE diagnostic temperature measurements. These profiles, fitted using a tension cubic spline, are shown in Fig 15. We do not present the raw HRTS data here, since these profiles are used below only to motivate a qualitative model.

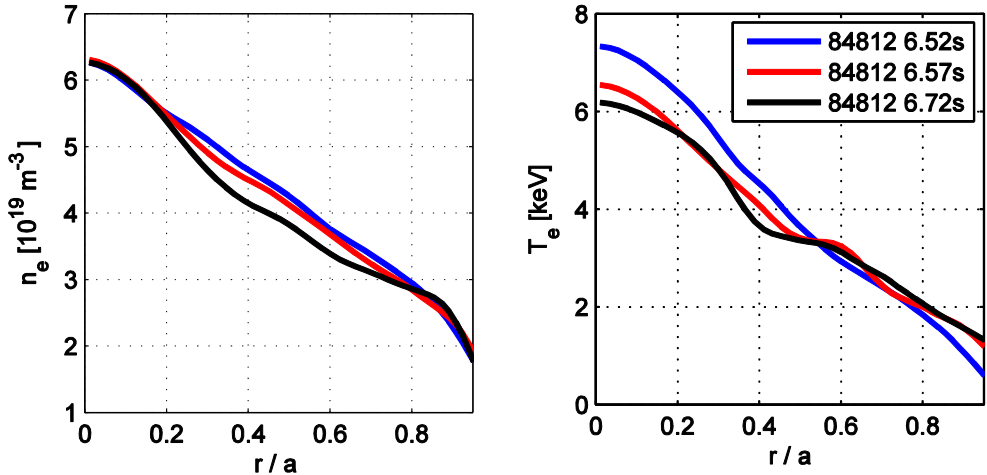

**Fig 15** *Fitted profiles of electron temperature ($T_e$) and density ($n_e$). t=6.52s is immediately before the n=2 mode appears and the later 2 times are after the n=2 mode forms.*

The ion temperature is not measured with enough fidelity to determine if it flattens in the island region, but is coupled to the electron temperature by collisional heat transfer (equipartition); here we assume $T_i = T_e$ for a first simple model. Flattening of the ion temperature inside an island has been observed in other devices (e.g. Fig 2a of Ref [37]).

The results show that turbulent W-transport after the island appearance (6.57s, Fig. 16b) is stabilised in the island region 0.5 < $r/a$ <0.6, (compare to 6.52s, Fig. 16a) due to the removal of the ion temperature gradient drive inside the island. The neoclassical W transport persists, but since the neoclassical diffusion is much smaller, steep density gradients can form (as in the inner core, see Fig 17). In addition, the neoclassical convection inside the island becomes strongly inward due to the reduction in temperature screening.

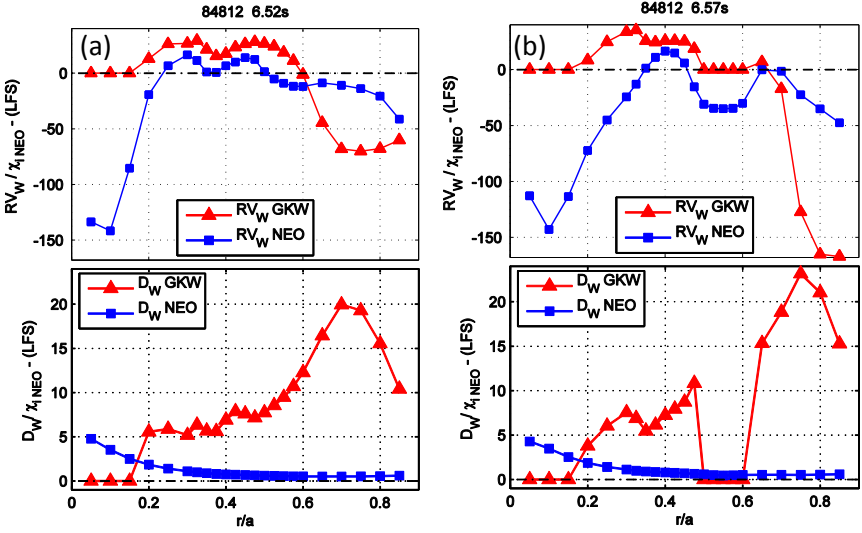

**Fig 16** *Predicted profiles of the W convection velocity (Upper row, negative values indicate inward convection), and diffusion coefficient (Lower row). The curves marked by triangles are from the gyro-kinetic (GKW) code and those marked with squares are from the neo-classical code (NEO) – the transport is combined additively. The left column, (a), is just before the n=2 mode and the right column, (b), is just after.*

Integrating the predicted gradients across the whole profile, the steady state model at 6.57s predicts significant increases in central W peaking due to these effects of the island (Fig 17). The magnitude of the W density (Fig 17 lower plot) is determined solely by the edge boundary value, which is not predicted by the model. The model predicts the logarithmic W-profile shape shown in Fig 17, but the magnitude is arbitrary and determined by the edge W-density. Once the W-density starts to peak on-axis, its evolution can undergo a non-linear feedback, where the associated increase radiation reduces temperature screening by affecting both the turbulence and increasing the density peaking inward impurity convection [11,37]. This can be seen in the later modelled time at 6.72s in Fig 17, in which the direct effect of the island has not changed, but the central W content has further increased, due to reduced temperature peaking and increased density peaking. The W profile predicted before the island forms can be more hollow, in quantitative agreement with the SXR measurements if the central $T_i$ profile is more peaked than $T_e$ (within measurement uncertainties). Irrespective of the profile inputs used, the qualitative effect of this island model is always to produce a strongly peaked W profile across the island, dramatically increasing the W content in the core.

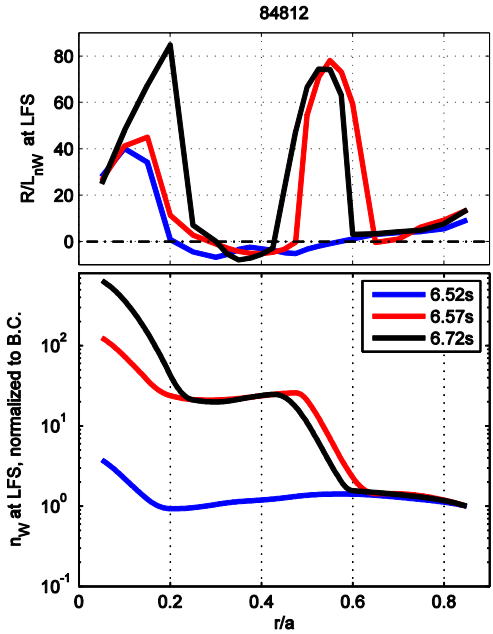

**Fig 17** *Upper plot; predicted W density peaking ($R/L_{nW}$) at the low field side. Lower plot; predicted steady state W density profile shape at the low field side just before the n=2 island (t=6.52s) and at 2 times just after the island forms.*

This model therefore qualitatively explains the observation that *n*=2 islands can accelerate the process of W accumulation in JET Hybrid plasmas, under the assumption that the magnetic island flattens the ion temperature profile but not the density profile, and has no direct effect on the W transport.

The model relies on a number of implicit assumptions, which we address here:

a) The island is treated in an axisymmetric geometry only through its effect on the 1D kinetic profiles, and the W transport models are not aware of any deviation from axisymmetric. However, since the W is in the Pfirsch-Schlüter collisional regime, deviations from axisymmetry do not have a large impact on its neoclassical transport [38]. The W ions do not feel the island through parallel transport, because the collisional timescale is shorter than the parallel transit time. Furthermore, W ions are deeply trapped due to the strong centrifugal force (with Mach numbers > 4, responsible for the clear LFS localisation seen in Fig. 11), and it is mainly passing particles that participate in parallel transport [39]. Thus the assumption of axisymmetry and the neglect of parallel transport of W ions around the island appear justified; more detailed modelling is in progress to validate this hypothesis.

b) Linear stability inside the island implies zero turbulent transport, which is probably an oversimplification. However, more complete non-linear gyro-kinetic simulations including a large imposed island have demonstrated substantially decreased (perpendicular) turbulent transport in the island region [40,41], so qualitatively our model should be valid.

c) The gradients of the deuterium density and temperature inside the island must be such that neoclassical convection is inward (approximately $R/L_{ni} > 0.5 R/L_{Ti}$). There is some evidence that this criterion should be satisfied [36, 37], which should be confirmed by more detailed measurements – or justified with more detailed theory and/or modelling. Given the persistence of a peaked density profile across the island, the internal flux surfaces the island cannot be considered analogous to the central flux surfaces of the tokamak: there is no expectation of W accumulation inside the island, since (unlike the tokamak centre) there is no reversal of density gradients across the island O-point, and so no reversal of neoclassical convection.

To summarise: the effect of the island on the fuel ion profiles, causes a significant reduction in the turbulence and decreases the neoclassical temperature screening, giving an unmitigated inward convection in the island region. The net effect is to cause a steep rise in the tungsten density from the outer to inner side of the $n=2$ island. This effect is irrespective of the shape of initial (pre $n=2$) tungsten density profile, whereas the model in section 5.1, in which the island enhances the tungsten diffusion directly, is sensitive to the relative location of the off-axis tungsten density peak and the island's radial location.

## 6. Conclusions and discussion

Hybrid plasma operation in JET, and possibly in ITER, offers a route to better confinement and higher $\beta$. However a significant range of MHD instabilities can occur in Hybrid operation. In JET the wall material was changed from carbon, to the JET-ILW combination of tungsten and beryllium. While this change in wall material has not significantly impacted the character of the MHD observed in Hybrid JET discharges, it has impacted the consequences. In JET-ILW operation it is observed that $n>1$ tearing modes can significantly speed the on-axis peaking of tungsten impurities, which in-turn degrades core confinement through radiative losses. Detailed examination shows the observed $n=2$ modes have a tearing character and are mainly driven by the plasma pressure (consistent with drive from the neo-classical bootstrap terms). Qualitative models for how low-$n$ islands can cause an initial off-axis impurity peak to rapidly peak on-axis, have been presented. If the island effect is to significantly enhance the tungsten diffusion then there is a strong sensitivity to the relative location of the initial off-axis tungsten peak and the island's radial location. If the island effect indirectly modifies tungsten transport, through its effect on the background fuel ion profiles (causing reduced diffusion and inward convection), then the indirect model, which has been built on physical first principles, shows less sensitivity. Further work is in progress to determine which model (direct or indirect) is dominant in accounting for the observed effects on tungsten transport. Mechanisms by which the island might directly affect the transport of the deeply trapped off-axis tungsten impurities need to be examined and the models need to account for the observed acceleration of tungsten accumulation with both $n=2$ and 3 modes, which have varying mode locations.

The most common instability is core $n=1$ activity in JET Hybrid pulses. The effects of this $n=1$ activity are mixed; sawteeth are beneficial in flushing out the core impurities, and also fishbone instabilities can decrease the core impurity level. Core $n=1$ activity can lower the central temperature (and gradients) and therefore through the indirect mechanism enhance central impurity accumulation. If $n=3$ instabilities occur, they are almost always during the phase of coherent core $n=1$ instability, and their combined effects can lead to significant confinement degradation.

It is known that ICRH can lead to outward convective velocities and give significant control over tungsten impurity accumulation – such beneficial effects have been observed in the baseline ELMy H-mode in JET [42] and further work is planned to exploit this effect in Hybrid plasmas on JET.

**Acknowledgements**

One of the authors (FJC) would like to thank Drs C. Angioni, J. García-Regaña, E. Poli, W.A. Hornsby, F. Parra Diaz and D. Zarzoso for useful discussions. This work has been carried out within the framework of the EUROfusion Consortium and has received funding from the Euratom research and training programme 2014-2018 under grant agreement No 633053 and from the RCUK Energy Programme [grant number EP/I501045]. To obtain further information on the data and models underlying this paper please contact